\documentclass[conference]{IEEEtran}
\IEEEoverridecommandlockouts

\usepackage{cite}
\usepackage{amsmath,amssymb,amsfonts,amsthm,thmtools}
\usepackage{algorithmic}
\usepackage{graphicx}
\usepackage{textcomp}
\usepackage{xcolor}

\usepackage{enumitem}
\usepackage{mathtools}
\usepackage[protrusion=false]{microtype}
\usepackage[ruled,linesnumbered,noend]{algorithm2e}
\usepackage[nocomma,short]{optidef} 
\usepackage{multirow}
\usepackage[para,flushleft]{threeparttable} 
\usepackage{subcaption}
\usepackage{graphbox}   
\usepackage{textcase}
\usepackage[hidelinks]{hyperref}
\usepackage[nameinlink]{cleveref}
\usepackage[super]{nth}
\usepackage{dblfloatfix}

\declaretheoremstyle[
    headfont=\bfseries,
    headindent = \parindent,
    postheadhook=\hspace*{\parindent},
    postheadspace = \newline
]{indentedbreak}

\declaretheorem[style=indentedbreak, name=Problem]{problem}

\def\BibTeX{{\rm B\kern-.05em{\sc i\kern-.025em b}\kern-.08em
    T\kern-.1667em\lower.7ex\hbox{E}\kern-.125emX}}

\makeatletter
\let\original@algocf@latexcaption\algocf@latexcaption
\long\def\algocf@latexcaption#1[#2]{%
  \@ifundefined{NR@gettitle}{%
    \def\@currentlabelname{#2}%
  }{%
    \NR@gettitle{#2}%
  }%
  \original@algocf@latexcaption{#1}[{#2}]%
}
\makeatother

\newcounter{solution}
\renewcommand{\thesolution}{\arabic{solution}}

\newcounter{SolutionLine}

\makeatletter
\newenvironment{solution}[1][htbp]
{%
  \let\oldalgocf\c@algocf
  \let\oldthealgocf\thealgocf
  \let\oldAlgoLine\c@AlgoLine
  \let\oldalgocfcurrentcounter\algocf@currentcounter

  \let\c@algocf\c@solution
  \renewcommand{\thealgocf}{\thesolution}

  \let\c@AlgoLine\c@SolutionLine
  \setcounter{SolutionLine}{0}

  \gdef\algocf@currentcounter{SolutionLine}

  \begingroup
  \crefalias{algocf}{solution}

  \begin{algorithm}[#1]
}
{%
  \end{algorithm}
  \endgroup

  \let\c@algocf\oldalgocf
  \let\thealgocf\oldthealgocf
  \let\c@AlgoLine\oldAlgoLine
  \let\algocf@currentcounter\oldalgocfcurrentcounter
}
\makeatother

\crefname{solution}{Solution}{Solutions}
\Crefname{solution}{Solution}{Solutions}

\crefname{step}{step}{steps}
\Crefname{step}{Step}{Steps}

\crefname{algocf}{alg.}{algs.}
\Crefname{algocf}{Algorithm}{Algorithms}

\crefname{problem}{problem}{problems}
\Crefname{problem}{Problem}{Problems}

\graphicspath{{figs/}}

\def\worktitle{High-Density Automated Valet Parking \\with Relocation-Free Sequential Operations}
\def\workname{DROP}
\def\worknamelong{high-\textbf{D}ensity \textbf{R}elocation-free sequential \textbf{OP}erations}
\def\workurl{\color{magenta}\url{https://drop-park.github.io}}

\newcommand{\projectsite}{\hyperlink{projectsite}{\color{magenta}project site}}

\begin{document}

\title{\worktitle}

\author{
    Bon Choe, Minhee Kang, Heejin Ahn
    \thanks{\scriptsize
        The Authors are with School of Electrical Engineering, Korea Advanced Institute of Science and Technology (KAIST), Daejeon, Republic of Korea. E-mails: \texttt{\{bonjae, ministop, heejin.ahn\}@kaist.ac.kr}

        This research was supported by the KAIST Convergence Research Institute Operation Program  and BK21 FOUR(Connected AI Education \& Research Program for Industry and Society Innovation, KAIST EE, No. 4120200113769). 
        }%
}

\maketitle

\begin{abstract}
In this paper, we present \workname, high-\textbf{D}ensity \textbf{R}elocation-free sequential \textbf{OP}erations in automated valet parking. \workname~addresses the challenges in high-density parking \& vehicle retrieval without relocations. Each challenge is handled by jointly providing area-efficient layouts and relocation-free parking \& exit sequences, considering accessibility with relocation-free sequential operations. To generate such sequences, relocation-free constraints are formulated as explicit logical conditions expressed in boolean variables. Recursive search strategies are employed to derive the logical conditions and enumerate relocation-free sequences under sequential constraints. We demonstrate the effectiveness of our framework through extensive simulations, showing its potential to significantly improve area utilization with relocation-free constraints.
We also examine its viability on an application problem with prescribed operation schedule.
The results from all experiments are available at: \hypertarget{projectsite}{\workurl}.

\end{abstract}


\section{Introduction}

Parking space is a critical bottleneck in fleet operations, especially in urban environments where available land is expensive and scarce. Automated valet parking systems (AVPS) enable autonomous vehicle maneuvering and support high-density parking configurations by reducing or eliminating driving aisles and buffer zones. High-density layouts have been shown to significantly improve area utilization~\cite{banzhaf2017survey}, making them attractive for increasing depot capacity within existing infrastructure.

However, high-density parking introduces operational challenges: in tightly packed layouts, vehicles obstruct one another, making relocation-free accessibility dependent on the current occupancy of surrounding spaces. We address the joint problem of layout efficiency and relocation-free sequential feasibility by generating high-density layouts and characterizing all parking and exit sequences executable without vehicle relocations.

\begin{figure}
    \centering
    \includegraphics[width=\columnwidth]{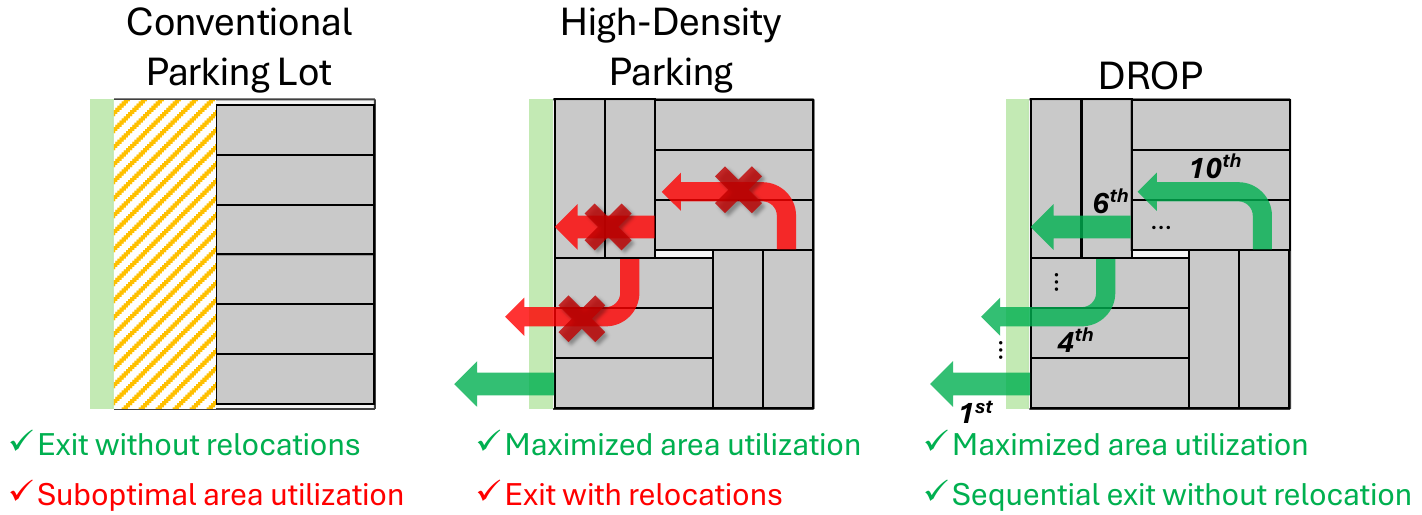}
    \caption{Comparison of the proposed framework \workname~with conventional parking lot and high-density parking.}
    \label{fig:concept}
\end{figure}

Prior work has addressed important parts of AVPS, but largely focused on either layout design or maneuver planning. High-density layout strategies such as k-Stacks~\cite{Timpner} and k-Deques~\cite{Banzhaf} improve storage capacity through structured patterns, while maneuver-centric~\cite{DOrey} and infrastructure-level~\cite{Nourinejad} studies address planning within fixed configurations. Puzzle-based parking~\cite{siddique2021puzzle} is most closely related, optimizing storage capacity and retrieval time on a discrete grid, but permits vehicle relocations and does not characterize relocation-free sequential feasibility. Consequently, no existing method systematically addresses this problem in high-density layouts.

We propose \workname~(\emph{\worknamelong}), which generates area-maximizing layouts and derives explicit boolean conditions for relocation-free accessibility, enabling enumeration of all valid sequences. The framework further identifies whether a prescribed operation schedule can be realized within this set.

The main contributions of this work are as follows:
\begin{itemize}
    \item We propose \workname, a framework for high-density AVPS for maximal area utilization together with relocation-free parking and exit sequences.

    \item We formulate relocation-free constraints as explicit logical conditions and employ recursive search to derive all valid sequences satisfying sequential constraints.

    \item We introduce an application problem that determines whether a prescribed operation schedule can be realized within the relocation-free sequence set derived by \workname.
    
    \item We validate \workname~through simulations across diverse parking lot dimensions, demonstrating high-density parking with complete relocation-free sequence enumeration.
\end{itemize}

\section{Problem Statement}\label{sec:ps}

In this section, we formally state our main problem of interest. We begin with motivation of our main problem, followed by defining the problem to be solved by \workname.

\subsection{Motivation}\label{subsec:moti}

Consider an AVPS depot operating in an unstructured parking lot. In pursuit of high area utilization, vehicles may be placed in a high-density layout without dedicated driving aisles and buffer zones. In such environment, the relocation-free accessibility of a vehicle is affected by not only the layout geometry, but also the vacancy states of obstructing vehicles. Additionally, vehicles in fleet operations may be parked and retrieved sequentially: vehicles are parked one by one during a parking phase, later retrieved one by one during an exit phase.
These observations motivate our \workname~framework that generates high-density layouts together with its relocation-free parking \& exit sequences.

Therefore, we state two core constraints: (1) relocation-free constraints and (2) sequential constraints. Based on following constraints, we present the problem definitions.

\begin{itemize}
    \item \textbf{Relocation-free constraints:} For each vehicle, there exists at least one kinematically feasible collision-free maneuver path between its allocated parking space and any entrance while only the vehicle moves and the other vehicles remain stationary (i.e., not relocated).
    \item \textbf{Sequential constraints:} Vehicles are parked and exited sequentially by two phases: a parking phase and an exit phase. First, vehicles are parked one by one during a dedicated parking phase until the lot reaches its capacity. Only after the parking phase is concluded does the exit phase begin, where vehicles are retrieved sequentially, one at a time. The exit phase is concluded when all parked vehicles are retrieved.
\end{itemize}

\subsection{Problem Definition}\label{subsec:definition}
We consider high-density parking problem with both relocation-free and sequential constraints. The problem is formally stated as below.

\begin{problem}[Layout Generation and Sequence Enumeration]\label{prob:1}
Given (1) a rectangular parking lot of width $W$ and length $L$, (2) an entrance set $\mathcal{E}$ of the parking lot and (3) identical vehicles with width $a$ and length $b$, find a set of parking stall arrangements that maximizes the number of parked vehicles, together with all relocation-free parking \& exit sequences.
\end{problem}

As an application of \Cref{prob:1}, we introduce a practical operational constraint introduced by operation schedule $\pi$. This is formulated as \textit{schedule mapping} between two executive days. Specifically, it is defined as one-to-one mapping between the arriving turn at the depot (parking sequence) and the departing turn from the depot (exit sequence).
This constraint requires that parked vehicles depart from the lot in a sequence dictated by $\pi$. With the schedule mapping and $N^*$ the maximum number of parked vehicles from the result of \Cref{prob:1}, we consider the problem of finding a pair of parking and exit sequences that comply with an operation schedule, as well as relocation-free and sequential constraints.

\begin{problem}[Application Problem with Operation Schedule]\label{prob:2}
Given $N^*$ vehicles and its operation schedule $\pi$, find a pair of parking and exit sequences satisfying both relocation-free and sequential constraints.
\end{problem}
To clarify, $\pi$ is a permutation of the vehicle set $\mathcal{V}$ with $|\mathcal{V}| = N^*$, i.e., a one-to-one mapping $\pi: \mathcal{V} \to \mathcal{V}$. Represented as a sequence
$
\pi = [\pi[0], \pi[1], \dots, \pi[N^*-1]],
$, the index $k$ indicates to the departure turn in the exit phase, and the value $\pi[k]$ identifies the specific vehicle by its arrival turn $k$. In this formulation, $\pi[k] = i$ implies that the vehicle arrived at $i$-th turn should depart at the $k$-th turn by $\pi$.

\section{Proposed Framework}\label{sec:framework}

\begin{figure}[!t]
    \centering
    \includegraphics[width=\linewidth]{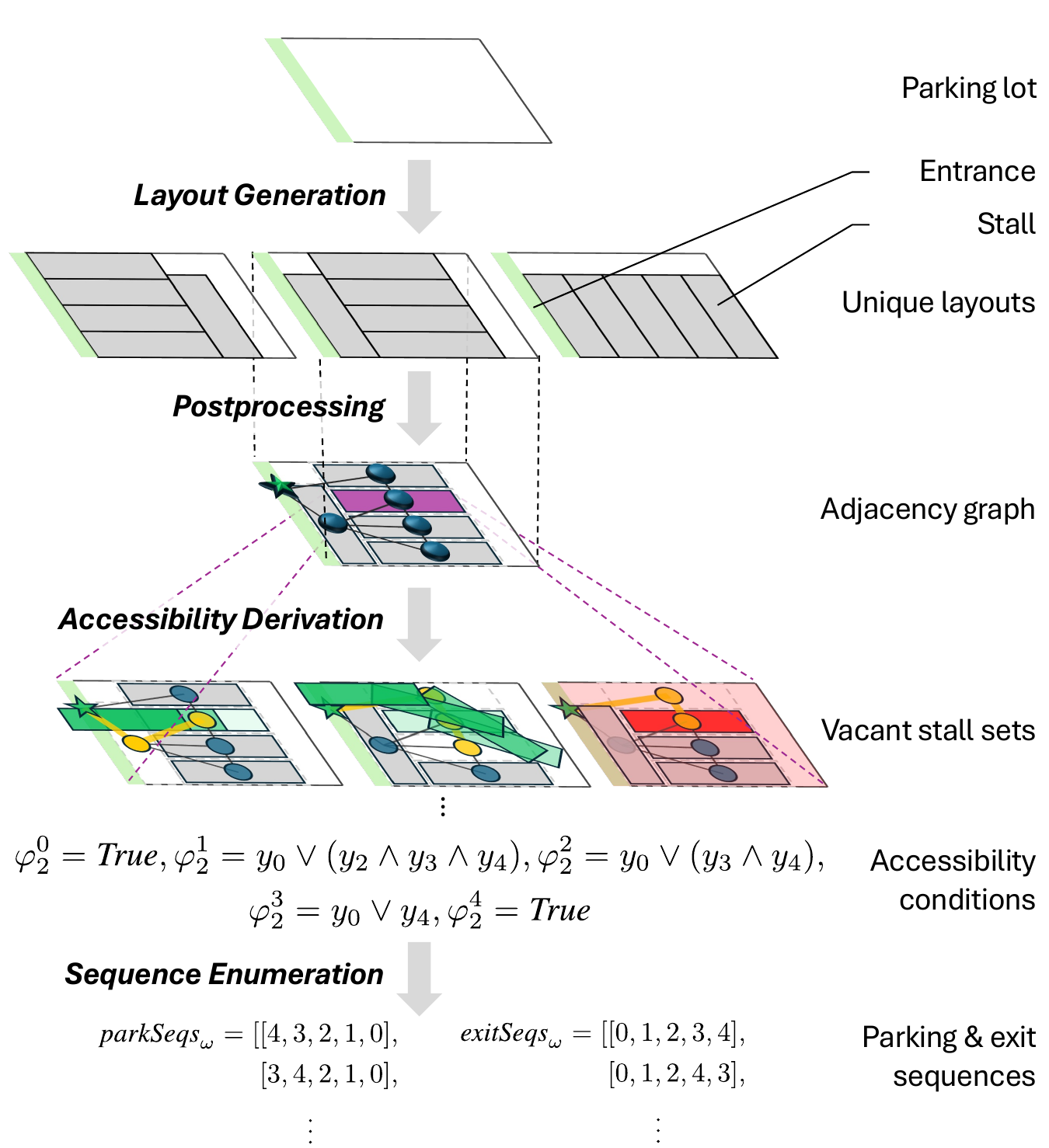}
    \caption{Overview of our \workname~framework.}
    \label{fig:overview}
\end{figure}

In this section, we describe \workname~to solve each problem. As illustrated in \Cref{fig:overview}, \textbf{\Cref{sol:1}} is designed to solve \Cref{prob:1} by generating unique layouts, deriving accessibility conditions and enumerating all relocation-free park \& exit sequences. Receiving outputs from \Cref{sol:1} and operation schedule $\pi$, \textbf{\Cref{sol:2}} is designed to solve \Cref{prob:2} by checking if a pair of relocation-free parking \& exiting sequences following operation schedule $\pi$ exists.

\subsection{Solution 1}\label{subsec:sol1}

\Cref{sol:1} takes as input the parking lot dimension (width $W$, length $L$), parking lot entrance set $\mathcal{E}$ and vehicle dimension (width $a$, length $b$). The output of this solution is a set of unique layouts $\Omega^*$, sets of all valid parking sequences $\{\textit{parkSeqs}_\omega\}_{\omega \in \Omega^*}$ and exit sequences $\{\textit{exitSeqs}_\omega\}_{\omega \in \Omega^*}$. The parking (or exit) sequence is defined as a sequence of parking stall indices, where the parking stall is stripped parking space in the parking layout.

To obtain the output, \Cref{sol:1} first generates unique layouts of the parking lot and postprocesses them by building adjacency graphs. Then it derives accessibility conditions $\varphi_\omega = \{\varphi_\omega^p\}_{p\in\mathcal{P}_\omega}$ for each feasible layout $\omega$ by recursively searching vacant stall sets for each stall $p$ in the stall set $\mathcal{P}_\omega$ in layout $\omega$. Based on the derived conditions, all valid parking and exit sequences $\textit{parkSeqs}_\omega, \textit{exitSeqs}_\omega$ are calculated following all accessible conditions.

\subsubsection{Layout Generation and Postprocessing}\label{subsec:layoutGen}

\begin{solution}[t]
    \caption{Pseudocode to solve \Cref{prob:1}}
    \label{sol:1}
    \KwIn{Parking lot dimension (width $W$, length $L$), \linebreak
    parking lot entrance set $\mathcal{E}$, \linebreak
    vehicle dimension (width $a$, length $b$)}
    \KwOut{A set of unique layouts $\Omega^*$,\linebreak
    a set of parking sequences $\{\textit{parkSeqs}_\omega\}_{\omega \in \Omega^*}$,\linebreak
    a set of exit sequences $\{\textit{exitSeqs}_\omega\}_{\omega \in \Omega^*}$}

    \textbf{Solve} \cite{silva2016pallet} to get a set of parking lot layouts $\Omega$\;\nllabel{line:solvePalletLoading}

    \textbf{Construct} a set $\Omega^*$ containing only topologically unique parking lot layouts from $\Omega$\;\nllabel{line:uniqueLayouts}

    \For{each parking lot layout $\omega \in \Omega^*$}{\nllabel{line:layoutLoopStart}
        \textbf{Build} Adjacency graph $\textit{Adj}(\omega,\mathcal{E})$\;\nllabel{line:buildAdjGraph}
        \lIf{layout $\omega$ is not feasible}{\textbf{continue}}\nllabel{line:infeasibleLayout}
        \For{each parking stall $p \in \mathcal{P}_\omega$}{\nllabel{line:stallLoopStart}
            \For{each entrance $e \in \mathcal{E}$}{
                $\textit{Adj}(\omega,e) \gets \textit{Adj}(\omega,\mathcal{E}) \setminus (\mathcal{E} \setminus \{e\})$\;
                $\varphi_\omega^p \gets \varphi_\omega^p \lor \textit{getAccessCond}(p, e, \textit{Adj}(\omega,e))$\;\nllabel{line:getAccesscond}
            }
        }\nllabel{line:stallLoopEnd}
        $(\textit{parkSeqs}_\omega, \textit{exitSeqs}_\omega) \gets \textit{getValidSequences}(\varphi_\omega)$\;\nllabel{line:getExitSeqs}

    }
    \Return{$(\Omega^*, \{\textit{parkSeqs}_\omega\}_{\omega \in \Omega^*}, \{\textit{exitSeqs}_\omega\}_{\omega \in \Omega^*})$}\;
\end{solution}

Layout generation adopts the pallet loading problem~\cite{silva2016pallet}, which maximizes the number of identical rectangular boxes placed within a larger rectangular pallet. We map the $W \times L$ parking lot to the \textit{pallet} and each $a \times b$ stall to the \textit{box}, allowing horizontal or vertical placement. Solving this problem yields a set of capacity-maximizing stall arrangements $\Omega$.
From $\Omega$, topologically unique layouts $\Omega^*$ are extracted by pushing all stalls to the bottom-left and retaining distinct configurations, reducing redundant computation.
For each $\omega \in \Omega^*$, an undirected adjacency graph $\textit{Adj}(\omega,\mathcal{E})$ is constructed with nodes representing stalls and entrances, and edges encoding geometric adjacency.

\subsubsection{Accessibility Analysis}\label{subsec:seqEnum}

To enumerate all valid parking \& exit sequences given a parking lot layout, an accessibility condition is defined where each parking stall in the layout satisfies a relocation-free constraint. Since a brute-force search is computationally intractable to find the set of the sequences, we devise the following three steps to reduce computational overhead: (1) skipping infeasible layouts, (2) deriving accessibility conditions for each stall in every feasible layout.

\paragraph{Skipping Infeasible Layouts}

\begin{algorithm}[t!]
    \caption{\textit{getAccessCond}($p, e, \textit{Adj}(\omega,e)$)}
    \label{alg:getAccessCond}
    \KwIn{Parking stall $p$, entrance $e$, adjacency graph $\textit{Adj}(\omega,e)$}
    \KwOut{Accessibility condition $\varphi$}

    $\textit{visitedSets} \gets \emptyset$\;
    $\varphi \gets \bot${\small\tcp*{Initialize with empty clauses}}

    \If{$\exists$ shortest path $q$ in $\textit{Adj}(\omega,e)$ from $p$ to $e$}{\nllabel{line:shortestPathInit}
        $\textit{expandConnectedSet}(p, e, \textit{Adj}(\omega,e), \text{Node}(q))$\; \nllabel{line:expandConnectedSet}
        $\textit{searchDetourPath}(p, e, \textit{Adj}(\omega,e), \text{Node}(q))$\;\nllabel{line:searchDetourPath}
    }
    
    \Return{$\varphi$}\;
    
    \SetKwProg{Fn}{Function}{:}{}
    \Fn{\textit{expandConnectedSet}($p, e, G, N_0$)}{ \nllabel{line:fnExpandConnectedSetStart}
        $\textit{deque} \gets \{N_0 \setminus \{p,e\}\}$\;
        \While{$\textit{deque} \neq \emptyset$}{
            $N \gets \textit{deque}.\text{pop\_front}()$\;
            \If{$N \in \textit{visitedSets}$}{\nllabel{line:pruneByVisited}
                \textbf{continue}\;
            }
            $\textit{visitedSets} \gets \textit{visitedSets} \cup \{N\}$\;
            \If{$\exists C \in \varphi$ such that $C \subseteq N$}{\nllabel{line:pruneByExistingClauses}
                \textbf{continue}\;
            }
            $\textit{obstacles} \gets \text{Node}(G) \setminus \{N \cup \{p, e\}\}$\;
            $\textit{waypoints} \gets \textit{HybridA*}(p, e, \textit{obstacles})$\;\nllabel{line:hybrid-astar}
            \If{$\textit{waypoints} \neq \emptyset$}{ \nllabel{line:ifFeasible}
                \lIf{$N \neq \emptyset$}{
                    $\varphi \gets \varphi \lor \bigwedge_{s \in N} y_s$
                }
                \lElse(\small\tcp*[f]{Always true}){$\varphi \gets \top$}
                \textbf{continue}\;
            }
                
            \If{$\textit{obstacles} \neq \emptyset$}{
                \For{each neighbor node $i$ of node set $N$ in the graph $G \setminus \{p, e\}$}{
                    $N^+ \gets N \cup \{i\}$\;
                    $\textit{deque}.\text{push\_back}(N^+)$\;
                }
            }
        }
    }\nllabel{line:fnExpandConnectedSetEnd}

    \Fn{\textit{searchDetourPath}($p, e, G, N_0$)}{\nllabel{line:fnSearchDetourPathStart}
        \For{each node $i \in N_0 \setminus \{p, e\}$}{
            \If{$\exists$ shortest path $q_i$ in $G \setminus \{i\}$ from $p$ to $e$}{
                $\textit{expandConnectedSet}(p, e, G, \text{Node}(q_i))$\;
                $\textit{searchDetourPath}(p, e, G \setminus \{i\}, \text{Node}(q_i))$\;
            }
        }
    }\nllabel{line:fnSearchDetourPathEnd}
\end{algorithm}

A layout is \textit{infeasible} if any stall has no kinematically feasible, collision-free path to any entrance regardless of vacancy states. We identify such layouts by evaluating each stall--entrance pair via Hybrid $A^*$ and discard them; for feasible layouts, a direct stall--entrance edge is retained in the adjacency graph.

\paragraph{Deriving Accessibility Conditions}\label{subsubsec:accessCond}
The accessibility condition $\varphi_\omega^p$ of stall $p$, defined in \Cref{eq:accessCond}, is a minimal DNF boolean expression over vacancy variables $y_s$, characterizing the minimal sets of other stalls that must be vacant for $p$ to be accessible without relocation, where $V = \text{Node}(\textit{Adj}(\omega,e))$, and $\textit{HybridA*}(p, e, \textit{obstacles})$ finds a kinematically feasible, collision-free path from $p$ to $e$ treating \textit{obstacles} as occupied stalls.

As shown in \Cref{alg:getAccessCond}, we employ an \textit{graph-based recursive search} to efficiently explore the space of stall sets in $\varphi_\omega^p$, as well as ensuring discovering all minimal clauses for the accessibility condition. The strategy includes (1) initializing with the shortest path, (2) recursively expanding search to adjacent stalls, and (3) exploring a detour path.

\begin{align}
    \varphi_\omega^p &= \bigvee_{e \in \mathcal{E}} \bigvee_{N \in \mathcal{S}_{p,e}^*} \bigwedge_{s \in N} y_s, \label{eq:accessCond} \\
    \mathcal{S}_{p,e}^* &= \{ N \in \mathcal{S}_{p,e} \mid \nexists N' \in \mathcal{S}_{p,e} \text{ s.t. } N' \subset N \} ,\label{eq:minimalSet} \\
    \mathcal{S}_{p,e} &= \{ N \mid N \cup \{p,e\} \text{ is connected in } \textit{Adj}(\omega,e), \nonumber \\
    &\quad \textit{HybridA*}(p, e, V \setminus (N \cup \{p,e\})) \neq \emptyset \nonumber \\
    &\quad \forall N \subseteq V \setminus \{p,e\} \} ,\label{eq:accessibleSet}
\end{align}

\begin{algorithm}[t]
    \caption{\textit{getValidSequences}($\varphi_\omega$)}
    \label{alg:getExitSeqs}

    \KwIn{Accessibility conditions $\varphi_\omega$}
    \KwOut{Set of all valid relocation-free parking and exit sequences ($\textit{parkSeqs}_\omega$, $\textit{exitSeqs}_\omega$)}

    $(\textit{exitSeqs}_\omega,\textit{currentSeq},\textit{exitedSet}) \gets (\emptyset,\emptyset,\emptyset)$\;
    $\textit{conds} \gets \varphi_\omega$\;

    \textit{enumerateDFS}(\textit{currentSeq}, \textit{conds}, \textit{exitedSet})\;
    $\textit{parkSeqs}_\omega \gets [\textit{reverse}(seq) \text{ for } seq \in \textit{exitSeqs}_\omega]$\;\label{line:parkSeqs}

    \Return{$(\textit{parkSeqs}_\omega$, $\textit{exitSeqs}_\omega)$}\;

    \SetKwProg{Fn}{Function}{:}{}
    \Fn{\textit{enumerateDFS}(\textit{seq}, \textit{conds}, \textit{exitedSet})}{
        \If{$|\textit{seq}| = |\mathcal{P}_\omega|$}{
            \textbf{Add} $\textit{seq}$ to $\textit{exitSeqs}_\omega$\;
            \Return\;\nllabel{line:completeSeq}
        }
        \If{$\nexists p$ s.t. $\textit{conds}[p]=\textit{true}$}{\nllabel{line:noTrueConds}
            \Return\;
        }
        
        \For {$p \in \mathcal{P}_\omega$}{
            \If{$p \notin$ \textit{exitedSet} and \textit{conds}$[p]$ = \textit{true}}{\nllabel{line:select-p}
                \textit{newSeq} $\gets$ \textit{seq}\;
                \textbf{Append} $p$ to \textit{newSeq}\;\nllabel{line:append-p}
                \textit{newConds} $\gets$ $\textit{conds}|_{y_p=\textit{true}}$\;\nllabel{line:updateConds}
                \textit{newExitedSet} $\gets \textit{exitedSet} \cup \{p\}$\;
                \textit{enumerateDFS}(\textit{newSeq}, \textit{newConds}, \textit{newExitedSet})\;
            }
        }

    }
\end{algorithm}

\begin{enumerate}[label=(\arabic*)]
    \item 
    \textbf{Initialization with Shortest Path} (\cref{line:shortestPathInit}):
    We initialize from the shortest path $q$ from $p$ to $e$ via Dijkstra's algorithm \cite{dijkstra}, as it yields a connected set of minimum size and thus the most compact initial clause. If no path exists, $p$ is inaccessible from $e$ and $\varphi$ returns $\bot$ (always \textit{false}).
    
    \item 
    \textbf{Connected set expansion} (\cref{line:expandConnectedSet}):
    Starting from $\text{Node}(q)$, we expand candidate connected sets via BFS \cite{clrs} using a double-ended queue (\textit{deque}) to process sets in order of cardinality. Each set $N$ is checked by Hybrid $A^*$ \cite{hybrid-astar}; if feasible, the corresponding conjunctive clause is added to $\varphi$. Two pruning rules limit evaluation: skipping already-visited sets and skipping supersets of existing clauses (\cref{line:pruneByVisited,line:pruneByExistingClauses}).

    \item
    \textbf{Detour path exploration} (\cref{line:searchDetourPath}):
    To discover alternative routes beyond the current path, we recursively explore detours (Function \textit{searchDetourPath}, \crefrange{line:fnSearchDetourPathStart}{line:fnSearchDetourPathEnd}). Specifically, for each intermediate node $i \in N_0 \setminus \{p,e\}$ on the current path, we temporarily remove $i$ and compute a shortest path $q_i$ from $p$ to $e$ in the reduced graph $G \setminus \{i\}$. If such a path exists, we pass its node set $\text{Node}(q_i)$ to \textit{expandConnectedSet} to continue the minimal-clause search, and then recursively call \textit{searchDetourPath} on $G \setminus \{i\}$ to further explore detours.
\end{enumerate}

\textbf{Complexity Analysis of \Cref{alg:getAccessCond}.} Let $n = |\mathcal{P}_\omega|$ denote the number of stalls and $M$ the maximum Hybrid A* iterations. In the worst case, \textit{expandConnectedSet} visits $O(2^n)$ node sets each at cost $O(M)$, and \textit{searchDetourPath} adds a factor of $O(n)$, yielding $O(2^n \cdot n \cdot M)$ per stall-entrance pair. In practice, the two pruning strategies substantially reduce Hybrid A* calls than the worst case, as many sets are either visited or subsumed by existing clauses, leading to significantly lower complexity.

\subsubsection{Sequence Enumeration}\label{subsubsec:seqEnumGenSeqs}

For each layout $\omega \in \Omega^*$, \Cref{alg:getExitSeqs} enumerates all valid exit sequences via DFS~\cite{clrs}, selecting at each step a stall $p$ with $\varphi_\omega^p = \textit{true}$ under the current vacancy state. The search terminates upon completing a full sequence or exhausting accessible stalls. Parking sequences are obtained by reversing each exit sequence (\cref{line:parkSeqs}).

\subsection{Solution 2}

\begin{solution}[!t]
    \caption{Pseudocode to solve \Cref{prob:2}}
    \label{sol:2}
    \KwIn{Parking lot layout $\omega$, \linebreak
    a set of all valid exit sequences $\textit{exitSeqs}_\omega$, \linebreak
    a set of all valid parking sequences $\textit{parkSeqs}_\omega$, \linebreak
    operation schedule $\pi$
    }
    \label{alg:pseudo2}
    \KwOut{A set of parking-exit sequence pair $\Sigma_\pi(\omega,\mathcal{E})$}
    $\Sigma_\pi(\omega,\mathcal{E}) \gets [(\textit{parkSeq}, \textit{exitSeq})$ for $\textit{parkSeq} \in \textit{parkSeqs}_\omega, \textit{exitSeq} \in \textit{exitSeqs}_\omega$ \linebreak if $\pi(\textit{parkSeq}) = \textit{exitSeq}$]\;\nllabel{line:getAssignments}
    \Return{\{$\Sigma_\pi(\omega,\mathcal{E})\}_{\omega \in \Omega^*}$}\;
\end{solution}

We now explain a \Cref{alg:pseudo2} to solve \Cref{prob:2}. Given an operation schedule $\pi$, we leverage valid relocation-free sequences $\textit{parkSeqs}_\omega$ and $\textit{exitSeqs}_\omega$ obtained in \Cref{subsec:sol1} and determine whether $\pi$ is compatible with layout $\omega$.

In particular, we filter parking-exit sequence pairs $(\textit{parkSeq}, \textit{exitSeq}) \in \textit{parkSeqs}_\omega \times \textit{exitSeqs}_\omega$ that follow operation schedule $\pi$. That is, the pairs should meet below condition:
\vspace{-1em}
\begin{align}
\pi(\textit{parkSeq}) &= [ \textit{parkSeq}[\pi[0]], \textit{parkSeq}[\pi[1]], \dots, \nonumber \\
                      &\quad \textit{parkSeq}[\pi[N-1]] ] \nonumber \\
                      &= \textit{exitSeq}\label{eq:filteringCondition}
\end{align}
\noindent where $\pi$ computes the constrained exit sequence from parking sequence and $N$ is a shorthand for the length of \textit{parkSeq} or \textit{exitSeq} (i.e., $N = |\textit{parkSeq}| = |\textit{exitSeq}|$). The pairs of parking and exit sequences satisfying \Cref{eq:filteringCondition} constitute $\Sigma_\pi(\omega,\mathcal{E})$.

\section{Simulation Experiment}\label{sec:experiment}

\subsection{Experiment Setup}

\begin{table}[t]
    \caption{Common Experiment Setup Configurations}
    \centering
    \label{tab:common-setup}
    \begin{tabular}{|l|l|c|}
        \hline
        \multicolumn{2}{|l|}{\textbf{Parameter}} & \textbf{Value} \\
        \hline\hline
        \multicolumn{2}{|l|}{Parking lot entrance} & Entire left boundary of the lot \\
        \hline
        \multicolumn{2}{|l|}{Parking stall size $(a \times b)$} & 3.0\,m $\times$ 9.5\,m \\
        \hline
        \multirow{5}{*}{Vehicle} & Size & 2.5\,m $\times$ 9.0\,m \\
        \cline{2-3}
        & Front/rear overhang & 1.885\,m / 2.875\,m \\
        \cline{2-3}
        & Wheelbase & 4.240\,m \\
        \cline{2-3}
        & Steering angle range & $[-\pi/6, \pi/6]$ \\
        \cline{2-3}
        & {Kinematic model} & Bicycle model \\
        \hline
        \multirow{3}{*}{Hybrid A*} & {Max. iterations} & $10^5$ \\
        \cline{2-3}
        & {Step size} & $[0.5\text{m}, 1\,\text{m}, 2\,\text{m}]$ \\
        \cline{2-3}
        & {Angle resolution} & $\pi / 36$ \\
        \hline
        \multicolumn{2}{|l|}{Operation schedules $\pi$} & All circular shifts of vehicle indices \\
        \hline
    \end{tabular}
\end{table}

The experiment environment mimics depot scenarios for 9\,m-length community buses in urban areas where parking space is limited. We evaluate three parking lot instances \texttt{15x12}, \texttt{20x16}, and \texttt{20x20}. All common configurations are summarized in \Cref{tab:common-setup}.

The parking lot entrance is set to the entire left boundary as a sole element of $\mathcal{E}$, imposing strict accessibility conditions. The stall size reserves a tight safety margin for buses with elongated wing mirrors, and vehicles are assumed parked at the center of each stall. The vehicle follows the bicycle model~\cite{bicycle}, and the operation schedule $\pi$ is evaluated over all circular shifts of vehicle indices.

Hybrid A*~\cite{hybrid-astar} is used as the path planner, with configuration parameters listed in \Cref{tab:common-setup}. The cost function $f = g + h$ penalizes traveled distance (with surcharges for reverse motion and direction switching) for $g$, and combines Euclidean distance with an orientation penalty for $h$.

The optimization problem in \cref{line:solvePalletLoading} is solved using Gurobi Optimizer~\cite{gurobi}. All experiments are implemented in C++ on a machine with 2 Intel Xeon Gold 6444Y 16-core CPUs and 8 64GB-RAMs, with OpenMP~\cite{openmp} parallelization applied to the iterative loops.

\subsection{Results}
In this paper, results (except computation time) are presented for the instance \texttt{15x12}, the one with the simplest and smallest parking lot among all tested instances. All results not shown in this paper are available on our \projectsite. Following the steps in \workname, intermediate results for each step are illustrated.

\subsubsection{Solution 1}
Results from each step are presented.

\paragraph{Layout Generation}

\begin{figure}[t!]
    \centering
    \includegraphics[width=\linewidth]{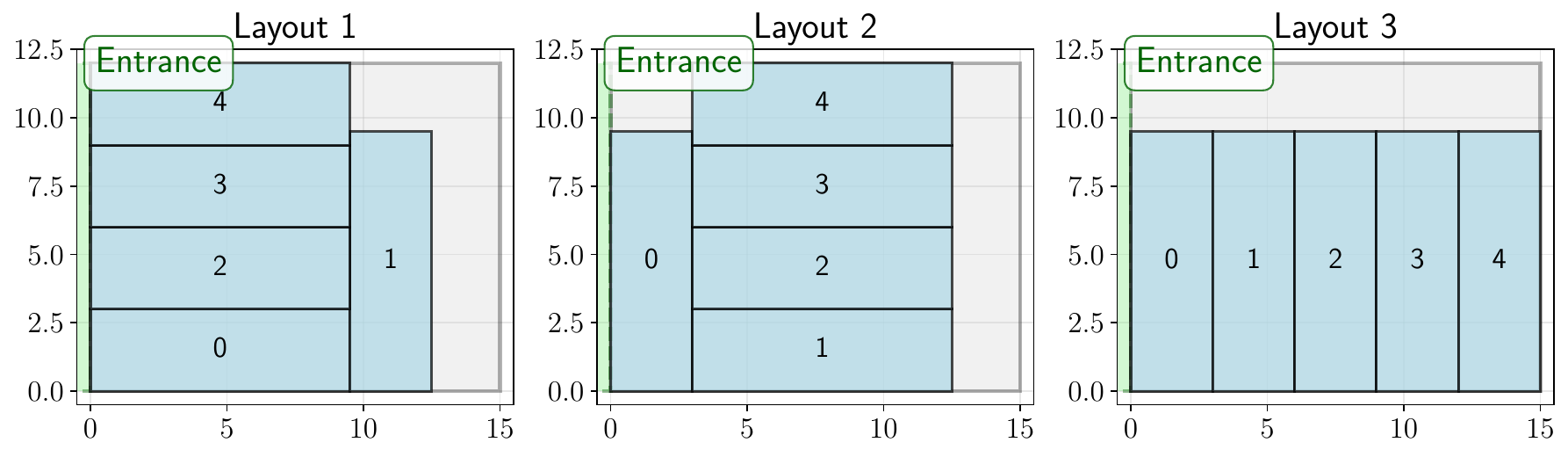}
    \caption{Unique parking lot layouts for the instance \texttt{15x12}.}
    \label{fig:unique-layouts}
\end{figure}

\begin{figure}[t!]
    \centering
    \includegraphics[width=\linewidth]{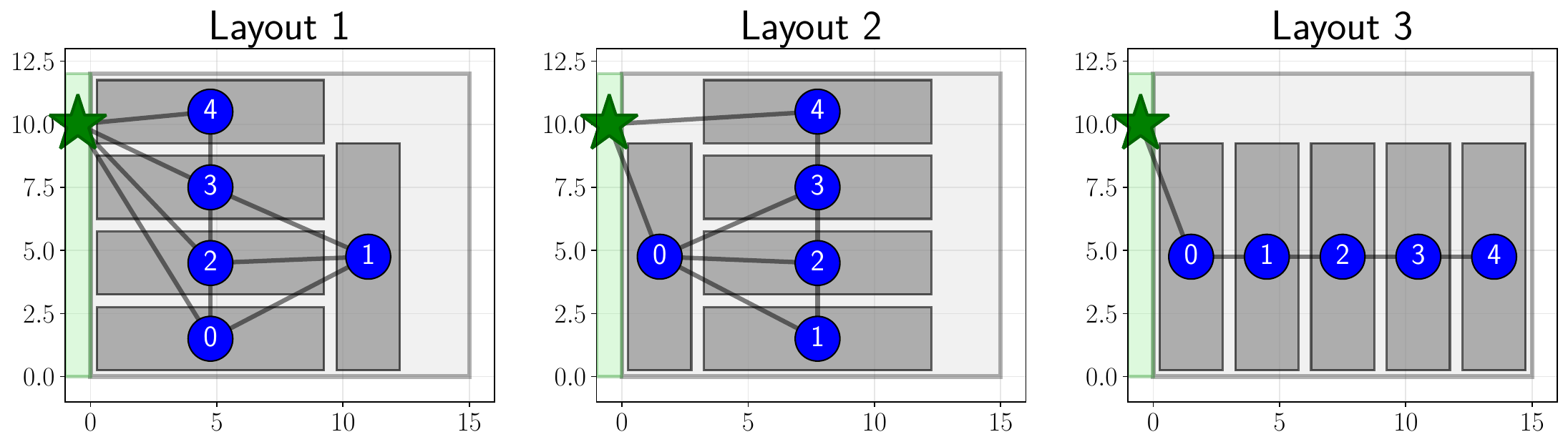}
    \caption{\small Adjacency graph for unique layouts for the instance \texttt{15x12}.}
    \label{fig:adj-graph}
\end{figure}

\begin{figure}[!t]
    \centering
    \begin{subfigure}[b]{0.3\linewidth}
        \centering
        \scalebox{0.7}{$\displaystyle
        \begin{aligned}
            \varphi_1^0 &= \textit{True} \\
            \varphi_1^1 &= (y_0 \land y_2) \lor (y_3 \land y_4)\\
            \varphi_1^2 &= \textit{True} \\
            \varphi_1^3 &= \textit{True} \\
            \varphi_1^4 &= \textit{True}
        \end{aligned}$}
        \caption{Layout 1}
    \end{subfigure}
    \begin{subfigure}[b]{0.3\linewidth}
        \centering
        \scalebox{0.7}{$\displaystyle
        \begin{aligned}
            \varphi_2^0 &= \textit{True} \\
            \varphi_2^1 &= y_0 \lor (y_2 \land y_3 \land y_4)\\
            \varphi_2^2 &= y_0 \lor (y_3 \land y_4) \\
            \varphi_2^3 &= y_0 \lor y_4 \\
            \varphi_2^4 &= \textit{True}
        \end{aligned}$}
        \caption{Layout 2}
    \end{subfigure}
    \begin{subfigure}[b]{0.3\linewidth}
        \centering
        \scalebox{0.7}{$\displaystyle
        \begin{aligned}
            \varphi_3^0 &= \textit{True} \\
            \varphi_3^1 &= y_0 \\
            \varphi_3^2 &= (y_0 \land y_1) \\
            \varphi_3^3 &= (y_0 \land y_1 \land y_2) \\
            \varphi_3^4 &= (y_0 \land y_1 \land y_2 \land y_3)
        \end{aligned}$}
        \caption{Layout 3}
    
    \end{subfigure}

    \caption{\small Accessibility conditions in each layout for the instance \texttt{15x12}.}\label{fig:accessibility-results}
\end{figure}

As a result from \Cref{sol:1}, three unique optimal parking lot layouts are obtained as \Cref{fig:unique-layouts}. For each obtained layout, an adjacency graph is constructed, as shown in \Cref{fig:adj-graph}. In the figure, stall and goal nodes are represented as blue circle and green star, respectively, and adjacent nodes are connected with edges.

\paragraph{Accessibility Analysis}
All three layouts pass the feasibility check, confirming at least one valid relocation-free exit sequence per layout. The derived accessibility conditions are shown in \Cref{fig:accessibility-results}. For instance, $\varphi_2^2 = y_0 \lor (y_3 \land y_4)$ indicates that stall~2 in Layout~2 is accessible without relocation if stall~0 is vacant, or if both stalls~3 and~4 are vacant.

\paragraph{Sequence Enumeration}

\begin{table}[h!]
    \caption{Relocation-free exit sequences of instance \texttt{15x12}}
    \label{tab:exit-sequence-results}
    \centering
    \begin{tabular}{cccc}
        \hline
        ~~~~~~~~~~~~~~~~~~~~~ & Layout 1 & Layout 2 & Layout 3 \\
        \hline
        $|\textit{exitSeqs}_\omega|$ & 56 & 34 & 1 \\
        \hline
    \end{tabular}
\end{table}

Based on the accessibility conditions in \Cref{fig:accessibility-results}, valid exit sequences are found for all layouts, summarized as \Cref{tab:exit-sequence-results}. Moreover, a set of all valid parking sequences $\textit{parkSeqs}_\omega$ are also obtained from $\textit{exitSeqs}_\omega$.

\paragraph{Computation Time}

\begin{table}[h!]
    \caption{Computation time comparison}
    \label{tab:computation-time}
    \centering
    \begin{tabular}{|c|l|cc|}
        \hline
        \textbf{Instance} & \textbf{Steps} & \textbf{Brute-Force} & \textbf{DROP} \\
        \hline
        \multirow{5}{*}{\texttt{15x12}}
            & Layout Gen.      & \multicolumn{2}{c|}{0.011 $\pm$ 0.003} \\
            & Postprocess      & --    & 0.0630 $\pm$ 0.0003 \\
            & Access. Analysis & --    & 0.3126 $\pm$ 0.0090 \\
            & Seq. Enum.       & 2.1961 $\pm$ 0.0791 & 0.0042 $\pm$ 0.0008 \\
            \cline{2-4}
            & Total            & 2.2071 $\pm$ 0.0790 & 0.3908 $\pm$ 0.021 \\
        \hline
        \multirow{5}{*}{\texttt{20x16}}
            & Layout Gen.      & \multicolumn{2}{c|}{0.026 $\pm$ 0.005} \\
            & Postprocess      & --    & 6.350 $\pm$ 0.099 \\
            & Access. Analysis & --    & 150.10 $\pm$ 0.630 \\
            & Seq. Enum.       & Intractable  & 2.561 $\pm$ 0.076 \\
            \cline{2-4}
            & Total            & Intractable  & 159.037 $\pm$ 0.822 \\
        \hline
        \multirow{5}{*}{\texttt{20x20}}
            & Layout Gen.      & \multicolumn{2}{c|}{0.053 $\pm$ 0.005} \\
            & Postprocess      & --    & 12.135 $\pm$ 0.236 \\
            & Access. Analysis & --    & 702.3 $\pm$ 15.1 \\
            & Seq. Enum.       & Intractable  & 505.3 $\pm$ 2.9 \\
            \cline{2-4}
            & Total            & Intractable  & 1219.788 $\pm$ 10.677 \\
        \hline
    \end{tabular}
\end{table}

\Cref{tab:computation-time} reports the wall-clock computation time (in s) for \Cref{sol:1} per instance, averaged over 10 independent runs. For comparison, we include a brute-force baseline that enumerates all $n!$ permutations and verifies relocation-free feasibility for each. \workname~achieves a 6$\times$ speedup over brute-force at \texttt{15x12}. For larger instances, brute-force becomes intractable as the $n!$ permutation space grows, whereas \workname~solves both instances within finite time.

\begin{figure*}[!t]
    \centering
    \includegraphics[width=.7\textwidth]{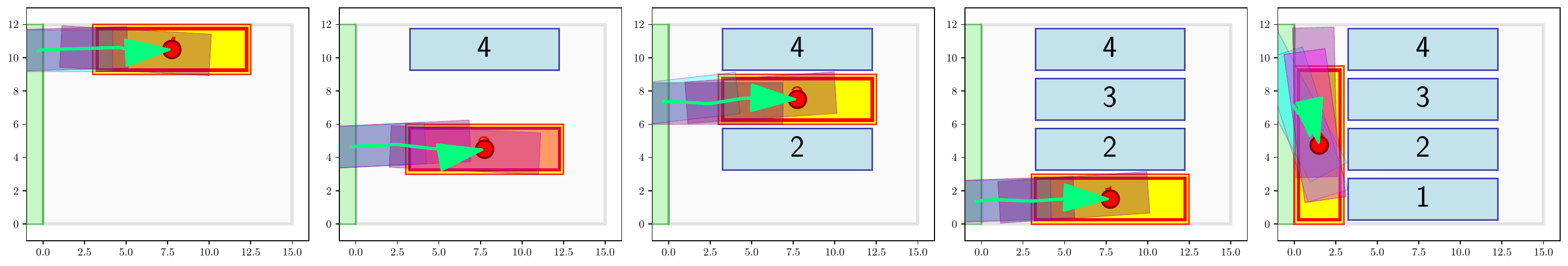}
    \includegraphics[width=.7\textwidth]{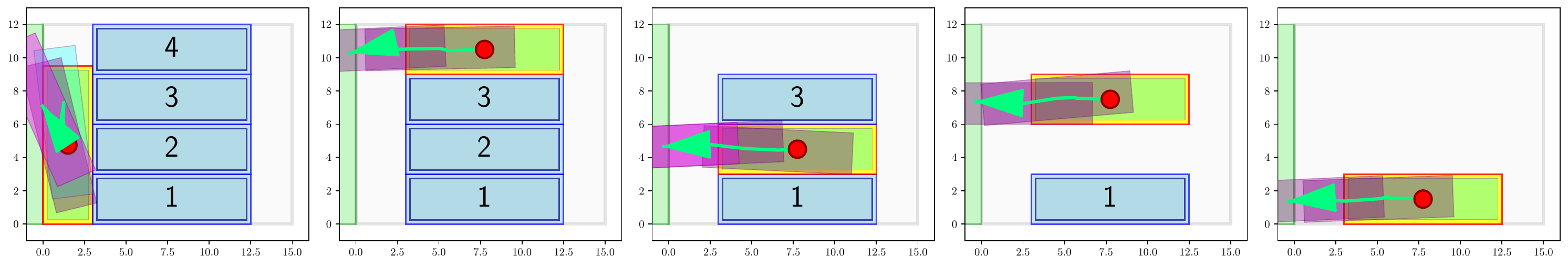}
    \caption{Example parking (\nth{1} row) and exit (\nth{2} row) sequences for layout~2 of \texttt{15x12} with $\pi = [4,0,1,2,3]$: $\textit{parkSeq} = [4,2,3,1,0]$ and $\textit{exitSeq} = [0,4,2,3,1]$. The current vehicle at each turn is colored in yellow (parking) and green (exiting).}
    \label{fig:final-allocations}   
\end{figure*}

\subsubsection{Solution 2}

\begin{table}[h!]
    \caption{The results from Solution 2 for instance \texttt{15x12}}
    \label{tab:allocation-results}
    \centering
    \begin{tabular}{cccc}
        \hline
        Operation Schedule $\pi$ & \textbf{Layout 1} & \textbf{Layout 2} & Layout 3 \\
        \hline
        $[0,1,2,3,4]$ &  \textbf{8} &  \textbf{2} & 0 \\
        $[1,2,3,4,0]$ & \textbf{24} &  \textbf{2} & 0 \\
        $[2,3,4,0,1]$ & \textbf{48} &  \textbf{4} & 0 \\
        $[3,4,0,1,2]$ & \textbf{40} & \textbf{12} & 0 \\
        $[4,0,1,2,3]$ & \textbf{16} & \textbf{26} & 0 \\
        \hline
        \multicolumn{4}{l}{%
        \begin{minipage}{6.5cm}%
            \vspace{.5em}
            \scriptsize Note: Each cell represents $|\Sigma_\pi(\omega,\mathcal{E})|$ for each layout $\omega$ and operation schedule $\pi$.
        \end{minipage}%
        }\\
    \end{tabular}
\end{table}

As shown in \Cref{tab:allocation-results}, Layouts~1 and~2 admit valid parking-exit sequence pairs for all cyclic operation schedules (highlighted in bold), while Layout~3 yields no valid pairs due to its single valid exit sequence. An example result for Layout~2 with $\pi = [4,0,1,2,3]$ is illustrated in \Cref{fig:final-allocations}: $\textit{parkSeq} = [4,2,3,1,0]$ and $\textit{exitSeq} = \pi(\textit{parkSeq}) = [0,4,2,3,1]$, a relocation-free pair conforming to the operation schedule. Each frame in \Cref{fig:final-allocations} shows the active vehicle with its trajectory, with stall indices labeled above each stall.

\subsection{Analysis on Parking Layouts and Relocation-free Sequences}
Optimized layouts reach stall coverage of approximately 79\%, 89\%, and 85\% for \texttt{15x12}, \texttt{20x16}, and \texttt{20x20}, respectively, while ensuring relocation-free retrieval sequences. Unlike structured high-density parking strategies that impose minimum lot dimension requirements, \workname~is applicable even to compact lots such as \texttt{15x12}.

As shown in \Cref{tab:exit-sequence-results}, the number of valid sequences for each layout increases as corresponding accessibility conditions (\Cref{fig:accessibility-results}) become less restrictive. A stall with a direct stall--entrance edge yields an always-true accessibility condition, meaning its vehicle can exit at any point regardless of surrounding vehicles. In \Cref{fig:adj-graph}, in Layout~1 of \texttt{15x12}, four of five stalls have direct stall--entrance edges, allowing many valid sequences than the other layouts; Layout~2 has two stalls with such edges, thereby reducing valid sequences than Layout~1.
By contrast, Layout~3 has only one such stall and its adjacency graph forms a chain: each stall's condition depends on all preceding stalls being vacant, creating a strict cascade. This yields only one valid sequence, which cannot satisfy any tested operation schedule (\Cref{tab:allocation-results}), making this layout unsuitable despite identical parking capacity.
These observations suggest that layouts with less constrained accessibility conditions yield larger, more flexible sequence set --- a useful guideline for layout design.

\section{Conclusions}\label{sec:conclusion}
\workname~was presented for high-density automated valet parking with relocation-free sequential operations. It generates capacity-maximizing layouts, derives relocation-free accessibility as explicit boolean conditions, and enumerates all valid parking and exit sequences. Given an operation schedule, \workname~further identifies schedule-consistent sequence pairs. Simulation results validate high area utilization around 80\% and above and relocation-free feasibility with respect to various operation schedules across multiple lot dimensions.
\workname~also supports heterogeneous fleets by incorporating each vehicle's dimensions and kinematic constraints, and is directly applicable to lots with multiple entrance points via additional entry edges in the adjacency graph.

Future directions include scalable heuristic strategies to address growing computation time according to lot size, formal theoretical guarantees on completeness and correctness, and extensions to settings with online departure requests.

\bibliographystyle{IEEEtran}
\bibliography{DROP}

\end{document}